\documentclass[12pt,a4paper]{article}
\usepackage{epsfig}

\title{Nonequilibrium effects due to charge fluctuations 
in intrinsic Josephson systems}
\author{D.A. Ryndyk, Institute for Physics of Microstructures, RAS, \\
GSP-105 Nizhny Novgorod 603600, Russia, \\ J. Keller, Institut f\"ur
Theoretische Physik, Universit\"at Regensburg, \\ D-93040 Regensburg, Germany,
\\ C. Helm, Los Alamos National Laboratory, T-11, \\ Los Alamos NM 87545, USA  }

\begin{document}

\maketitle

\begin{abstract}

Nonequilibrium effects in 
layered superconductors forming a stack of  intrinsic Josephson
junctions are investigated. We discuss two basic nonequilibrium 
effects caused by charge fluctuations on the superconducting layers: 
a) the shift of the chemical potential of the condensate and 
b) charge imbalance of quasi-particles, and study their influence on IV-curves
and the position of Shapiro steps. 

Pacs: 74.72.-h, 74.50.+r, 74.40.+k. 
\end{abstract}

\section{Introduction}

For the strongly anisotropic cuprate superconductors
Bi$_2$Sr$_2$CaCu$_2$O$_{8+\delta}$ \newline (BSCCO) and
Tl$_2$Ba$_2$Ca$_2$Cu$_3$O$_{10+\delta}$ (TBCCO) the
electromagnetic transport perpendicular to the CuO$_2$ layers
can well be described by a model where the CuO$_2$ planes with the
intermediate material form a stack of Josephson junctions. This intrinsic
Josephson effect can be seen in the multibranch structure of the IV-curves but
also in the behavior of the material in external magnetic fields and under
high frequency irradiation \cite{Kleiner92,Kleiner94,Yurgens96, Yurgens00}.
Here also Shapiro steps have been observed
\cite{Kleiner94,Doh00}. 

In the presence of a bias current each junction of the stack can be either in
the resistive or superconducting state. In the resisistive state a finite 
dc-voltage $V$ appears together with voltage oscillations
with a frequency $\omega$ given by the
Josephson relation $\hbar \omega =2e V$. Both effects are 
accompanied with static and oscillating charge
fluctuations on the layers. In a system of weakly coupled very thin atomic
layers such charge fluctuations may lead to nonequilibrium effects
described by a shift of the chemical potential of the condensate and a 
charge imbalance between electron and hole-like quasi-particles
\cite{Koyama96,Bulaevskii96a, Artemenko97,Ryndyk98,Preis98,Ryndyk99,Helm01}.
This is different in classical
Josephson systems with massive superconducting electrodes. Here similar charge
fluctuations occur, however, they are restricted to the surface of the
superconducting electrode and do not affect the superconductor as a whole and
in particular do not reach the normal electrode. 

\begin{figure}
\epsfig{figure=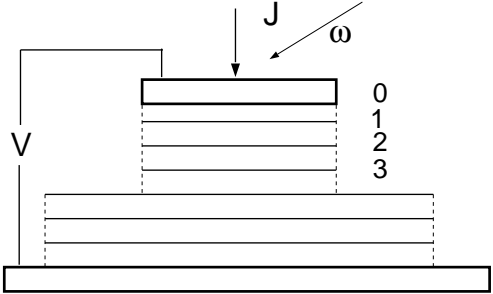,height=4cm}
\caption{Typical mesa-structure used for 2-point measurements consisting of
a stack of superconducting layers $n=1,2 \dots$, a normal electrode $n=0$ on
top and a base
electrode. \label{fig.1}}
\end{figure}

It is a non-trivial question wether such nonequilibrium effects
have measurable consequences. In an experiment like that of Clarke
\cite{Clarke72}  several
contacts are necessary to measure the difference between the chemical
potential of condensate and quasi-particles.  Experiments on cuprate
superconductors are mostly done on mesa structures (see figure 1), where the top
layer is covered with one gold electrode, and the same electrode is used
to inject the current and  measure the voltage (for a 4-probe measurement see
\cite{Yurgens96}). Also in such situations 
nonequilibrium  effects should be observable  for instance through their
influence on the IV-curves: there should be a
difference in the total voltage depending on wether two resistive junctions
are neighbors in the stack or separated by a junction in the superconducting
state. Irregularities in the brushlike structure of the IV-curves, however,
can also be produced by other effects, i.g. fluctuations in the critical
current density. Therefore in this paper we investigate in particular the
possible influence of nonequilibrium  effects on the voltage-position
of Shapiro steps, which should be more precise. We also add some theoretical
considerations for measurements on systems with multiple contacts, where it is
possible to measure the charge imbalance relaxation directly. 

In a Shapiro-type  experiment high-frequency radiation is applied to the sample, in
this case to the antenna connected to the gold contact on top of the mesa.
For  sufficiently strong radiation the internal Josephson oscillation
frequency of a junction in the resistive state  is locked to the
external frequency $\omega$ fixing the voltage in a small current interval.
For a single junction in the resistive state one expects a set of Shapiro
steps in the IV-curve at voltages $V= V_{\rm cont}+ m\hbar \omega/(2e)$. Here
$V_{\rm cont}$ is the voltage due to the contact resistance between the normal
electrode and the first superconducting layer. It is determined by
measuring separately the IV-curve in the superconducting state. A similar
result is expected, if m junctions are in the resistive state and we restrict
ourselves to the first Shapiro step in each branch. 
In this paper we will show
that the position of steps may be changed into
$V= V_{\rm cont}+ m\hbar \omega/(2e)-\delta V$, where again $V_{\rm cont}$
is the contact resistance in the superconducting state. This happens if
the first resistive junction is close to the normal electrode. The apparent
reduction of the Shapiro step voltage is indirectly due to a change of the
contact  voltage in the presence of a nonequilibrium charge distribution on
the first superconducting layer, and $\delta V$ is proportional to the
relaxation time of the quasi-particle charge. A similar result is obtained for
the position of the m$^{\rm th}$ Shapiro step on the first branch. 

In a recent experiment on a BSCCO sample a downshift of the first Shapiro step
on the first branch has been observed \cite{Rother01}. Experiments on
other samples show more complicated structures which could not be interpreted
completely. Therefore in this paper we concentrate
on the theoretical aspects of nonequilibrium effects in layered
superconductors. In the following section we summarize the basic results of a
theory developed by one of us (D.A.R.) \cite{Ryndyk99}  which is based on a
kinetic  equation for the distribution function describing charge imbalance.
First this theory is used to study the influence
of nonequilibrium effects on the IV-curves. Then these results
are applied to the calculation of Shapiro steps and finally to the
analysis of systems with multiple contacts. In all these applications of the
nonequilibrium theory we restrict ourselves to stationary effects. We also
neglect the influence of high frequency oscillations on the
dc-component of the supercurrent, which is a reasonable approximation for 
weakly coupled superconducting layers (large values of the McCumber
parameter). Some results from  the numerical simulation of the
time-dependent kinetic equations can be found in \cite{Ryndyk99}.

Let us briefly mention here also the relation of the present theory to other
theoretical approaches. As mentioned above there are
two nonequilibrium effects connected with charge fluctuations a) the shift of
the chemical potential of the condensate, b) the charge imbalance of
quasi-particles (there is also an influence of nonequilibrium effects on 
the amplitude of the superconducting order parameter, which will not be
considered here). If we take into account the shift of the chemical
potential but neglect charge imbalance we essentially obtain the results of the
theory developed by Koyama and Tachiki \cite{Koyama96}. In the approach by Artemenko
and Kobelkov \cite{Artemenko97} and by our group \cite{Preis98, Helm01}  a
systematic perturbation theory in the gauge-invariant scalar potential is
used. In these theories charge imbalance is considered only indirectly as far
as it is induced by fluctuations of the gauge invariant scalar potential. The 
present theory is more general. Here charge imbalance is taken into account as
an indepent degree of freedom and therefore the results are
different from earlier  treatments. A detailed comparison of the different
approaches  will be given in a future publication.

 \section{Theory
of stationary nonequilibrium effects in layered superconductors}

\subsection{Generalized Josephson relation}

We consider a system of
superconducting layers $n$ with superconducting order parameter $\Delta_n(t) =
\vert \Delta \vert \exp(i\chi_n(t))$ with time-dependent phase $\chi_n(t)$
neglecting a possible time dependence of the amplitude of the order parameter.
The basic quantity which enters all the Josephson relations is the 
gauge invariant phase difference  between layer $n$ and
$n+1$ given by  
\begin{equation} 
\gamma_{n,n+1}(t)= \chi_n(t) -
\chi_{n+1}(t) - \frac{2e}{\hbar} \int_n^{n+1} dz A_z(z,t) , 
\end{equation}
where $A_z(z,t)$ is the vector potential in the barrier. For the time
derivative of $\gamma_{n,n+1}$ we obtain the generalized Josephson relation:
\begin{equation} 
\dot \gamma_{n,n+1} = \frac{2e}{\hbar} \Bigl( V_{n,n+1} +
\Phi_{n+1} - \Phi_n\Bigr) .
\end{equation} 
Here  
\begin{equation}
V_{n,n+1}= \int_n^{n+1}dz E_z(z,t)
\end{equation}
is the voltage  and $\Phi_n(t)$ is the so-called 
gauge invariant scalar potential defined by 
\begin{equation}
\Phi_n(t)= \phi_n(t) - \frac{\hbar}{2e} \dot \chi_n(t),  
\end{equation}
where $\phi_n(t)$ is the electrical scalar potential. In this paper $e$
denotes the elementary charge. The charge of the electron is $-e$.
The quantity $\hbar \dot \gamma_{n,n+1}$ is the total energy to transfer a 
Cooper pair between neighboring layers $n$ and $n+1$. The quantity
$e\Phi_n$ can be considered as shift of the chemical potential of the
superconducting condensate with respect to an avarage chemical potential. It
determines the minimum in k-space of the quasi-particle excitation
energy. 

In the following we have to consider fluctuations of the charge densities on
the superconducting layers. For this purpose let us briefly mention
some basic  concepts for the description of nonequilbrium superconductors as
introduced by Tinkham and Clarke \cite{Clarke72}. In the BCS theory one may
split the total electronic charge density on each   layer into $\rho = \rho^s +
\rho^q$ with  $\rho^s = \frac{2}{Vol}\sum_{k} v_k^2$ and $\rho^q
=\frac{2}{Vol}\sum_k (u_k^2 - v_k^2) f_k$. In equilibrium the latter vanishes
because there are equal numbers of electron-like  and hole-like excitations.
In a nonequilibrium state of the superconductor a finite quasi-particle
charge may exist, which is called charge imbalance. The quantity $\rho^s$,
sometimes  called condensate charge, should not be confused
with the  superfluid charge density, which vanishes at the transition temperature. 
On the contrary $\rho^s$ 
approaches the total density in the normal state. Fluctuations of the two
charge densities  are caused not only by  shifts in the chemical potential
entering the quasi-particle excitation energy and the functions $u_k, v_k$,
but also by fluctuations in the distribution functions $f_k$. In a
quantum-kinetic theory based on quasi-classical Green's functions the same
distinction between the two types of charge densities can be made. The
k-dependent functions  are then replaced by corresponding energy-dependent
quantities. 

A fluctuation in the (2-dimensional) charge density $\rho^s_n$ is directly
related to the shift of the chemical potential in the layer $n$:
\begin{equation}  
\rho_n^s(t)= - 2e^2 N(0) \Phi_n(t) .
\end{equation} 
Here $N(0)$ is the 2-dimensional density of states of the
conduction electrons in the layer. Unlike a true chemical potential, which is
defined only in equilibrium, the quantity $\Phi_n(t)$ may be time dependent.
In the following we will, however,  restrict ourselves 
to stationary processes where $\Phi_n$ is independent of time.

It is convenient to express also fluctuations in the charge imbalance with help
of a quasi-particle potential $\Psi_n$ by writing formally 
\begin{equation}
\rho^q_n = 2e^2N(0)\Psi_n(t).
\end{equation}
Then we obtain for the total charge density fluctuation:
\begin{equation}
\rho_n(t) = -2e^2N(0)(\Phi_n(t) -\Psi_n(t)) . \label{rho}
\end{equation}
The calculation of the charge imbalance $\rho^q_n$ will be discussed in section
(2.3) for the stationary case. 

With help of (\ref{rho}) and the Maxwell equation  ($d$ is
the distance between the layers) 
\begin{equation}
\rho_n= \frac{\epsilon_0}{d} (V_{n,n+1}-V_{n-1,n}) \label{Maxwell}
\end{equation}
the generalized Josephson relation now  reads:
\begin{equation}
\frac{\hbar}{2e }\dot \gamma_{n,n+1} = (1+2 \alpha)
V_{n,n+1} - \alpha (V_{n-1,n} + V_{n+1,n+2}) + \Psi_{n+1} - \Psi_n
\label{genJos}
\end{equation}
with $\alpha= \epsilon_0/(2e^2N(0)d)$. It shows that the Josephson
oscillation frequency is determined not only by the voltage  in the same
junction but also by the voltages  in neighboring junctions. Furthermore it is
influenced by the quasi-particle potential $\Psi$ on the
layers. If we neglect the latter we obtain for $\dot \gamma_{n,n+1}$ the same
result as in \cite{Koyama96}. 

For a Josephson junction in the presence of sufficiently strong 
high-frequency irradiation 
the average Josephson oscillation
frequency is locked to the external frequency $\omega$. From the
generalized Josephson relation (\ref{genJos}) we obtain for the
m$^{\rm th}$ Shapiro step
\begin{eqnarray} 
m\frac{\hbar  \omega}{2e} &=&
\frac{\hbar}{2e }\langle \dot \gamma_{n,n+1}\rangle 
\\
&=& (1+2\alpha) V_{n,n+1} - \alpha (V_{n-1,n}+V_{n+1,n+2}) + \Psi_{n+1} -
\Psi_n, \nonumber
\label{Shapiro}
\end{eqnarray}
where the dc-components of the quantities on the r.h.s have to be used. 
This relation can also be used to describe the voltage for a barrier in the
superconducting state, then the l.h.s. is zero:
\begin{equation} 
0= (1+2\alpha) V_{n,n+1} - \alpha (V_{n-1,n}+V_{n+1,n+2}) + \Psi_{n+1} -
\Psi_n.
\label{Shapiro0}
\end{equation}

\subsection{Current equation}

In order to calculate the electronic transport in layered superconductors we
need expressions for the current and charge density. 
The results are derived with help of a Keldish technique
for nonequilibrium Green's functions. In particular the quasi-particle charge
is calculated from a kinetic equation for the corresponding quasi-particle 
distribution function. For the current density the following approximate
result is obtained:
\begin{equation}
j_{n,n+1}= j_c \sin\gamma_{n,n+1} + \frac{\sigma_{n,n+1}}{d} \Bigl(
\frac{\hbar}{2e} \dot \gamma_{n,n+1} + \Psi_n-\Psi_{n+1}\Bigr) .\label{current}
\end{equation}
The first term is the current density of Cooper pairs. The rest is the
quasi-particle current density (we neglect here the interference term
proportional to $\cos \gamma$). The physical meaning of the latter
becomes more apparent if we 
express these terms with help of (\ref{rho},\ref{Maxwell}) by the charge
densities: 
\begin{equation} 
j_{n,n+1}= j_c \sin\gamma_{n,n+1} +
\frac{\sigma_{n,n+1}}{d} \Bigl( V_{n,n+1} + \frac{\rho_n}{2e^2N(0)}
-\frac{\rho_{n+1}}{2e^2N(0)}\Bigr) , \label{current1}  
\end{equation} 
which can also be written as:
\begin{equation} 
j_{n,n+1}= j_c \sin\gamma_{n,n+1} +
\frac{\sigma_{n,n+1}}{d} \Bigl((1+2\alpha) V_{n,n+1} -\alpha(V_{n-1,n} +
V_{n+1.n+2}) \Bigr) . \label{current2}  
\end{equation} 
Then we see that the quasi-particle current is driven not only by the voltage
but also by a diffusion term proportional to the charge difference in the two
layers. 
The expression (\ref{current1}) for the current density  is valid
for temperatures close to $T_c$. For $T \ll T_c$
the conductivities are different for the two contributions of the quasi-particle
current (they correspond to the quantities $\sigma_0$ and $\sigma_1$ introduced
by Artemenko et al. \cite{Artemenko97}). For simplicity we neglect this
difference in the following. 

Note that (\ref{current1}, \ref{current2}) stay meaningful also in the
normal state for a system of junctions with different conductivities. Then
$\alpha$ describes the shift of the chemical potential due to charge
accumulation in the normal state.

In the stationary state the current density $j_{n,n+1}$ is
the same for all barriers and can be replaced by the bias current density $j$.
If the Josephson junction between layers $n$  and $n+1$ is in the resistive
state (without high-frequency irradiation), then we may use the current relation
(\ref{current2}) with a vanishing dc-component of the supercurrent (which 
is a good  approximation in the limit of large values of the McCumber
parameter) and obtain:
\begin{equation} 
\frac{j d}{\sigma_{n,n+1}} = 
(1+2\alpha) V_{n,n+1} -\alpha(V_{n-1,n} +
V_{n+1,n+2}). \label{current3}   
\end{equation} 

For a correct determination of the total voltage of a stack as
shown in figure 1 an investigation of the influence of nonequilibrium effects on
the contact voltage between the normal electrode and the first superconducting
layer is necessary.  The physical nature of this contact is not known
precisely. It may vary between a tunneling contact and a metallic contact with
superconductivity partially suppressed on the first superconducting layer by
proximity effect. In the following we assume that this contact can be
described by a tunneling contact between a thick normal electrode and a thin
superconducting layer. Then we may neglect the shift of the chemical potential
in the normal layer and obtain for the current density between the normal
electrode ($n=0$) and the first superconducting layer ($n=1$) (compare with
(\ref{current3})) 
\begin{equation}  
\frac{j d}{\sigma_{0,1}}=
(1+\alpha)V_{0,1} - \alpha V_{1,2}  . \label{contact}
\end{equation}  
There might be a small shift of the chemical potential on
a thin layer at the surface of the normal electrode. This can be included
in the theory, but has no influence on the total voltage.
For notational simplicity we have assumed here that the barrier width $d$
between the normal electrode and the first superconducting layer is the same
as that between two superconducting layers. This can easily be generalized
but does not change the content of the final results.
Note that the quasi-particle potentials appear 
explicitly only in the equations derived from the 
generalized Josephson equation and not in the current equations 
(\ref{current3}, \ref{contact}).

\subsection{Relaxation of the quasi-particle charge}

In addition to the current equation the microscopic theory provides us with 
an equation for the charge density. As the quasi-particle charge is only
part of the total density an additional mechanism for the charge imbalance
relaxation has to be introduced. In the stationary case the following result
for the quasi-particle charge is obtained \cite{Ryndyk99}:
\begin{equation}
\rho_n^q= -2e^2N(0) \Psi_n=\tau_q (j_c \sin\gamma_{n,n+1} - j_c
\sin\gamma_{n-1,n}) ,  \label{qprelax}
\end{equation}
where $\tau_q$ is the charge imbalance relaxation time. 

This result can be made plausible in a simple way: Let us start
from a balance equation for the superfluid charge density:
\begin{equation}
0 = j_c\sin\gamma_{n-1,n} - j_c \sin\gamma_{n,n+1} - 
 \frac{\partial \rho_n^q}{\partial t}\vert_{conv} . \label{scont}
\end{equation}
The first two terms on the r.h.s describe the change of the superfluid density due to
supercurrents between the neighboring layers. In the stationary state this is
balanced by the conversion of quasi-particle charge into  condensate charge
(the last term in (\ref{scont})). We assume that in the stationary case 
this conversion can be described by a relaxation process  
\begin{equation}  
\frac{\partial \rho_n^q}{\partial t}\vert_{conv} = - \frac{1}{\tau_q} \rho_n^q . 
\label{relprocess}
\end{equation}
Using this epression in (\ref{scont}) the
result (\ref{qprelax}) is obtained.

The relation (\ref{qprelax}) can also be written in another way which will be useful in
the following. Using the continuity equation for the total current density in
the stationary case we find for the quasi-particle current density:  
\begin{equation}
0 = j^{qp}_{n-1,n} - j^{qp}_{n,n+1} + 
 \frac{\partial \rho_n^q}{\partial t}\vert_{conv} , \label{qprelax1} 
\end{equation}
then 
\begin{equation}
\Psi_n= d(j^{qp}_{n-1,n} - j^{qp}_{n,n+1})/\sigma_q
\label{qprelax2} 
\end{equation}
with $\sigma_q = 2e^2_0N(0)d/\tau_q$. 

For the later discussion it will be useful to introduce a relaxation
parameter $\eta_{n,n+1} \equiv \sigma_{n,n+1}/\sigma_q$.
In the following we will assume that nonequilibrium effects are small, i.e.
$\alpha\ll 1$, $\eta_{n,n+1} \ll 1$. In first order in $\alpha$ and $\eta$ 
we may then use the approximation 
$j^{qp}_{n,n+1}= \sigma_{n,n+1}V_{n,n+1}/d$ for the calculation of $\Psi_n$ 
and obtain
\begin{equation}
\Psi_n \simeq 
\eta_{n-1,n} V_{n-1,n} - \eta_{n,n+1}V_{n,n+1} ,
\label{qprelax3} 
\end{equation}
or for a barrier in the resistive state we may replace  $j^{qp}_{n,n+1}$ in
(\ref{qprelax2}) by the total current density $j$. 

The relaxation parameter $\eta$ can also be
written as $\eta = \tau_q/\tau_t$, where $\tau_t$ is a typical tunneling time. 
In the case of weak nonequlibrium, $\eta \ll
1$ considered here, the
charge imbalance has relaxed before quasi-particles have tunneled to the next
layer.   

The charge imbalance potential $\Psi_n$ vanishes for a layer between
two junctions in the superconducting state, when the current is carried
exclusively by Cooper pairs. It vanishes also for a layer between two junctions
in the resistive state, when the current is carried primarily by
quasi-particles. A non-zero value of $\Psi_n$ can be found, in particular, for
a layer between one junction in the superconducting and one junction in the
resistive state.  As $\tau_q$ is proportional to $1/\Delta$ close to $T_c$, but
$j_c$ is proportional to $\Delta^2$ the charge imbalance vanishes at $T_c$.

These results are derived here for the stationary case, which will be
needed for the discussion of dc-properties in the following. In the
presence of a time-dependent chemical potential $\Psi_n(t)$ an additional 
term proportional to $d\Psi_n(t)/dt$ has to be added in
(\ref{relprocess}). The results for the time-dependent case are discussed in 
\cite{Ryndyk99}.

\section{Influence of nonequilibrium effects on the IV-curves }

As a first application of the theory outlined above we study the influence of
nonequilibrium effects on the IV-curves, i.e. we calculate the total voltage
for a stack of junctions, where one or more Josephson junctions are in the
resistive state in the absence of  high-frequency irradiation. Special
attention is paid to the contact voltage. Here we consider explicitly only one
contact. The effect of the other contact can be added easily to the final
result.   

 \subsection{All junctions in the superconducting state}

Let us start the discussion with the case where all junctions are in the
superconducting state. Then the voltages are determined by the following set
of equations: Equation (\ref{contact}) is used for the contact with the normal electrode
and (\ref{Shapiro0}) with $n\ge 1$  for the junctions in the
superconducting state. 
Adding up these equations we obtain for the total voltage 
which is the contact voltage in the superconducting
state: 
\begin{equation}
V= V_{\rm cont}= \frac{jd}{\sigma_{0,1}} + \Psi_1 .
\end{equation}
For the quasi-particle potentials  we have:
\begin{equation}
\Psi_1= \eta_{0,1} jd/\sigma_{0,1}= jd/\sigma_q, \quad \Psi_n=0 \quad {\rm for}
\quad n \ge 2. 
\end{equation}
Then the contact voltage is 
\begin{equation}
V_{\rm cont}= V^0_{\rm cont}+ \delta V 
\end{equation}
with $\delta V = jd/\sigma_q$ and the "bare" contact voltage $V^0_{\rm cont}
=jd/\sigma_{0,1}$.
We see that the contact voltage
is changed by the charge imbalance induced by the quasi-particle current on
the first superconducting layer.

\subsection{One or two  junctions in the resistive state}

If the first Josephson junction (between layers 1 and 2) is in the resistive
state, but all the other junctions are in the superconducting state, 
then the contact with the normal electrode is again
described by (\ref{contact}). For the resistive junction we use 
the current relation 
(\ref{current3}) with $n=1$, for the superconducting junctions 
(\ref{Shapiro0}) with $n \ge 2$.

Adding up these equations we find for the total voltage:
\begin{equation}
V= \frac{jd}{\sigma_{0,1}} + \frac{jd}{\sigma_{1,2}} + \Psi_2 , 
\end{equation}
where we have  used $\Psi_n=0$ for $n \ge  3$. 
The nonequilibrium effect now comes from the second layer, 
where $\Psi_2= \eta_{1,2} jd/\sigma_{1,2}=jd/\sigma_q$. Here and in the
following we assume that the charge imbalance relaxation rate
is the same on all superconducting layers. 
Then the total voltage can also be written as
\begin{equation}   
V= V_{\rm cont} + \frac{jd}{\sigma_{1,2}} . 
\end{equation}
The IV-curve  of
the first branch appears at the usual voltage after subtracting the
contact voltage. This is different, if the Josephson junction in the
resistive state is not close to the normal contact. Then one obtains for the
total voltage: 
\begin{equation}
V=V_{\rm cont}+ \frac{jd}{\sigma_{n,n+1}}( 1+ 2\eta_{n,n+1})= 
V_{\rm cont} + \frac{jd}{\sigma_{n,n+1}} + \delta V
\end{equation}
with $\delta V=2jd/\sigma_q$. The first branch has a slightly larger 
voltage for the same current
density. The voltage shift $\delta V$ is due to the charge imbalance potentials
generated by the quasi-particle currents on the two layers $n$ and $n+1$
between the resistive junction and the neighboring superconducting junctions. 

Of special interest is the case where  two junctions are  in the resistive
state, if we compare the total voltage for the case  where the resistive
junctions are neighbors with the situation where they are separated by
one ore more  barriers in the superconducting state. In the latter case
the total voltage is again larger by $\delta V= 2jd/\sigma_q$. This shift is
due to the charge imbalance potential on the two additional boundaries  
separating the resistive junctions from the
junctions in the superconducting state. This result can be easily
generalized to the case of several resistive junctions. 

In the present theory
the shift of the chemical potential described by the coefficient $\alpha$ has
no influence on  the total voltage. This  is a consequence of the special form
of the quasi-particle current density used here (compare with
(\ref{current}, \ref{current2})). This result is  different from earlier
treatments \cite{Matsumoto99}, where for the quasi-particle current density a simple 
ansatz of the form $j^{qp}_{n,n+1} = \sigma V_{n,n+1}/d$ has been used.

\section{Influence on Shapiro steps}

For a Josephson junction in the presence of strong high-frequency irradiation the
current relation (\ref{current3}) is not useful, since on a Shapiro step the
current is not fixed by the voltage. Instead the Josephson oscillation
frequency is locked to the external radiation. Then we have to use the generalized Josephson relation 
equation (\ref{Shapiro}) to determine the voltage. Let us assume that the first
Josephson junction (between layers 1 and 2) is in the resistive state and 
all other junctions are in the superconducting
state,  then the voltages for the different  barriers  
are again determined by (\ref{contact}, \ref{Shapiro0}), but for the
resistive junction with a first Shapiro step we have to use 
\begin{equation}
\frac{\hbar\omega}{2e}= (1+2\alpha)V_{1,2} - \alpha(V_{0,1}+V_{2,3}) +
\Psi_2-\Psi_1 .
\end{equation}  
With $\Psi_n=0$ for $n\ge 3$ we obtain for the total voltage:
\begin{equation}
V= \frac{jd}{\sigma_{0,1}}  + \frac{\hbar\omega}{2e} + \Psi_1 .
\end{equation}
In this case the quasi-particle potential $\Psi_1$  is given by
\begin{equation}
\Psi_1= d( j^{qp}_{0,1} - j^{qp}_{1,2})/\sigma_q .
\end{equation}
As $j^{qp}_{0,1}=j$ and $j^{qp}_{1,2}\simeq j$ at the center of the Shapiro
step, $\Psi_1$ vanishes in this case. 
Therefore we find for the
total voltage 
\begin{equation}
V = V^0_{\rm cont} + \frac{\hbar \omega}{2e} , 
\end{equation}
where $ V^0_{\rm cont} =jd/\sigma_{0,1}$ is the "bare" contact voltage. If we
introduce instead the contact voltage which is measured in the
superconducting state  we have:
\begin{equation} 
V = V_{\rm cont} + \frac{\hbar \omega}{2e} - \delta V 
\end{equation} 
with $\delta V = jd \eta_{0,1}/\sigma_{0,1} = jd/\sigma_q$.  

The apparent shift of the Shapiro step for the first Josephson junction is
due to the difference of the charge imbalance potential $\Psi_1$ on the first
superconducting layer in the pure superconducting state and in the resistive
state (the effect of the charge imbalance $\Psi_2$ on the second layer drops
out in the total voltage). 
In order to get such an experimental  shift of the Shapiro step voltage it 
is essential that the
Josephson junction  in resonance with the external
radiation is close to the normal electrode. Otherwise the quasi-particle
potentials $\Psi_n$ drop out in the total voltage. 

The result can easily be extended to the case where the first $m$ Josephson
junctions are in the resistive state and in resonance with the external
radiation. In that case we find 
\begin{equation}
V= V_{\rm cont} + m \frac{\hbar\omega}{2e} 
       -\delta V 
\end{equation}
with $\delta V = jd \eta_{0,1}/\sigma_{0,1} = jd/\sigma_q$.  
Again it is assumed that the first Josephson junction in
the resistive state is close to the normal electrode, otherwise $\delta
V=0$. Note that only the voltage of the Shapiro step on the first branch is
shifted, the difference between the other Shapiro steps is still at the "right"
voltage $\hbar\omega/(2e)$. 

A similar result is obtained for higher order Shapiro steps on
the first branch. Then the voltage of the m$^{\rm th}$ Shapiro step is given by
\begin{equation}
V = V_{\rm cont}(j) + m\frac{\hbar \omega}{2e} - \delta V(j) 
=V^0_{\rm cont}(j) + m\frac{\hbar \omega}{2e} . 
\end{equation} 
As the Shapiro steps appear here at different current values we have to take
into account the current-dependence of the contact voltage explicitly.
Then all Shapiro steps appear to be shifted if the contact voltage in the
superconducting state is subtracted.

\section{4-point measurements}

In the preceding section we have investigated the influence of
charge imbalance on the shift of Shapiro steps. This effect is indirectly due
to a change of the  contact voltage between the normal electrode and the first
superconducting layer caused by charge imbalance. With help of a 4-point
measurement it is possible to determine the contact voltage separately.
Furthermore a direct determination of the charge imbalance and its
relaxation rate will be possible. 

We consider a stack with two normal electrodes on top (see figure 2). 
Then  one contact can be used to inject a
current and the other contact to measure the
voltage in the absence of a current. 

\begin{center}
\begin{figure}
\epsfig{figure=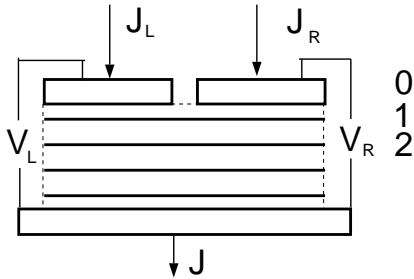,height=4cm}
\caption{Stack with two normal electrodes used to measure the
charge imbalance relaxation rate.  \label{fig.2}}
\end{figure}
\end{center}

Let us denote the current densities and voltages of the left and right
electrode (with length $l_{L,R}$) by 
$j_{L,R}$ and $V_{L,R}$. We assume that the currents injected through the
left and right electrode are different, in particular we are interested in
the case $j_R=0$.

First we consider the case that all inner junctions
are in the superconducting state. Then the individual voltages are determined
by the following set of equations: 
\begin{equation} 
\frac{j_{L,R} d}{\sigma_{0,1}}= (1+\alpha)V^{L,R}_{0,1} - \alpha V^{L,R}_{1,2}
,
\end{equation} 
\begin{equation}
0= (1+2\alpha) V^{L,R}_{n,n+1} - \alpha (V^{L,R}_{n-1,n} +
V^{L,R}_{n+1,n+2}) + \Psi_{n+1} - \Psi_n,\quad {\rm for} \quad n \ge 1 .
\end{equation}
The injected currents induce a charge imbalance potential  $\Psi_1$ on the
first superconducting layer controlled by the
quasi-particle currents.  We assume that the diffusion length of
quasi-particles along the layer is large compared to the width of the stack.
Then  we obtain a uniform charge imbalance potential: 
\begin {equation}
\Psi_1= (\frac{l_L}{l} j_L +  \frac{l_R}{l} 
j_R)\frac{d}{\sigma_q} . 
\end{equation}
Note that the total charge on the first superconducting layer in general is not
uniform and the voltages $V_L$ and $V_R$ are different, but the
electrochemical potential which is controlled by  $\langle \dot
\gamma_{n,n+1}\rangle $ is constant along the layers. 

For the voltages at the left and right contact we then obtain:
\begin{equation} 
V_{L,R} = \frac{j_{L,R} d}{\sigma_{0,1}} + \Psi_1 .
\end{equation}
In the case $j_R=0$ we find in particular:
\begin{equation}
V_L - V_R= \frac{j_Ld}{\sigma_{0,1}}, \quad V_R =
V_L \frac{l_L}{l} \eta_{0,1} .
\end{equation}
which allows us to measure the relaxation parameter $\eta_{0,1}$ and the
"bare" contact voltage $j_Ld/\sigma_{0,1}$ separately.

Now we consider the case that the barrier between the first two
superconducting layers is in the resistive state. Then a similar calculation
leads to 
\begin{equation} 
V_{L,R} = \frac{j_{L,R} d}{\sigma_{0,1}} + \frac{jd}{\sigma_{1,2}} + \Psi_2
\end{equation}
with $\Psi_2= jd \eta_{1,2}/\sigma_{1,2}= jd/\sigma_q$. Now the nonequilibrium
effect comes from the second superconducting layer. Note that the current
density flowing through the barrier (1,2) is the total current density $j$.
In the special case $j_R=0$ a comparison between $V_L$ and $V_R$ allows us to
determine the "bare" contact  voltage of the left electrode. 

Finally let us discuss the results for applied high-frequency radiation. If the
barrier between the first two superconducting layers is in the  resistive
state and is in resonance with the radiation frequency $\omega$ the first
Shapiro step appears at the voltage
\begin{equation} 
V_{L,R} = \frac{j_{L,R} d}{\sigma_{0,1}} +
\frac{\hbar\omega}{2e} + \Psi_1 
\end{equation}
with $\Psi_1\simeq 0$ in the center of the Shapiro step. In particular, for
$j_R=0$ the voltage at the right electrode measures the Shapiro step voltage
directly without the contact voltage. The same is true for the voltage
of the m$^{\rm th}$ Shapiro step on the first branch. 

\section{Summary}

In this paper we have discussed the influence of nonequilibrium effects on
stationary properties of Josephson contacts in layered superconductors. In
particular, we have studied the influence on the IV-characteristics and
on  the voltage-position of Shapiro steps. We find that from the two basic
nonequilibrium effects, a) shift of the chemical potential of the condensate
and b) charge imbalance of quasi-particle charges, only the latter has an
influence on these properties, while the chemical potential shift drops out in
the total voltage. This result depends crucially on the form of the 
quasi-particle current density and the correct treatment of the contact
resistance. Note that for time-dependent processes, like collective modes, 
also the coupling between superconductng layers due to the chemical potential 
shift is relevant \cite{Koyama96}.

In our theory the charge imbalance relaxation plays an important role.
Microscopically this relaxation is caused by inelastic
scattering processes within the layer. In the case of d-wave superconductors
also elastic scattering may contribute. In the present treatment a
phenomenological relaxation rate has been introduced.

In our investigation of Shapiro steps we find a shift of the voltage of the
first Shapiro step on the first resistive branch of the IV-curve if the 
Josephson junction in the resistive state is close to the normal electrode.
This apparent shift, which is not a violation of the basic Josephson
relations, can be traced back to a change of the contact voltage
between the normal electrode and the first superconducting layer due to
charge imbalance on the first superconducting layer in the presence of a finite
quasi-particle current. The distance between higher order Shapiro steps on the
first branch is also changed due to the current dependence of the contact
voltage. 

Recently Shapiro step experiments have been performed in the THz regime on
mesa structures of BSCCO \cite{Rother01}. The voltage-position of the Shapiro
step has been determined by subtracting the contact voltage in the
superconducting state in the absence of radiation. In one experiment a  3\%
downshift of the step position on the first branch has been
observed, i.e. the ratio between charge imbalance lifetime and tunneling
time is $\eta= \tau_q/\tau_t= 0.03$. In other samples also Shapiro steps on higher branches
could be detected. Their positions showed a more complicated structure which
up to now is not fully understood. Therefore it is not yet possible to 
present a detailed comparison with experimental data. 

The difficulty with the proper subtraction of the contact voltage can be avoided in  4-point
measurements. Therefore we have investigated the influence of charge
imbalance on the Josephson effect for an experimental set-up where the first
superconducting layer is in contact with  two normal electrodes, and only
one electrode is used to inject the current. Also in this case the contact
voltage is influenced by the charge imbalance produced by the quasi-particle
currents, but the contact voltage can be determined separately by comparing the
voltages measured at the two electrodes. Taking that into account, the 
Shapiro steps appear again at the right position.

\section{Acknowledgement}

We would like to thank P. M\"uller and his group for a very
fruitful cooperation on the investigation of Josephson effects in
high-temperature superconductors. Their  experiments have initiated the
present theoretical investigation on nonequilibrium effects. This work has
been supported by the Bayerische Forschungsstiftung in a common project.
D.A. Ryndyk would like to thank the Graduiertenkolleg
"Nonlinearity and nonequilibrium effects in condensed matter" for a post-doc
followship supported by the German Science Foundation.  C. Helm gratefully acknowledges
financial support by the US Department of Energy under contract W-7405-ENG-36.

\end{document}